\documentclass[%
 reprint,
 amsmath,amssymb,
 aps,
]{revtex4-2}

\usepackage{graphicx}% Include figure files
\usepackage{dcolumn}% Align table columns on decimal point
\usepackage{bm}% bold math
\usepackage{hyperref}% add hypertext capabilities
\usepackage{cancel}
\usepackage[normalem]{ulem} % [normalem] prevents it from changing \emph to underline
\usepackage{xcolor}

\definecolor{nicered}{rgb}{0.7,0.1,0.1}
\definecolor{nicegreen}{rgb}{0.1,0.5,0.1}

\begin{document}

\title{Axions on a Hyperbolic Ride:\\ Geometric Suppression of CMB Isocurvature and a Blue-Tilted Spectrum}

\author{Sai Chaitanya Tadepalli}
\email{saictade@iu.edu}
\affiliation{Physics Department, Indiana University, Bloomington, IN 47405, USA}

\begin{abstract}
CMB limits on cold-dark-matter isocurvature are often interpreted as excluding
the simultaneous realization of high-scale inflation and large QCD axion decay
constants in pre-inflationary Peccei--Quinn (PQ) scenarios. This conclusion can be evaded by exploiting \emph{field-space geometry}. For a minimal complex PQ scalar with a $U(1)$-symmetric potential and a nonlinear sigma-model kinetic term $d\sigma^{2}=dR^{2}+f^{2}(R)\,d\theta^{2}$, the observable axion fluctuation is $\delta\theta\sim H_{\rm inf}/f(R)$, so an enhanced effective decay constant $f(R)$ suppresses isocurvature without explicit PQ breaking, extreme radial displacements, or additional couplings. We specialize to a hyperbolic metric $f(R)\propto \sinh(R/L)$ with curvature scale $L$. The same geometry also induces a time-dependent $\mathcal{O}(H_{\rm inf})$ effective mass  for the canonical axial mode during radial slow-roll, and fixing the tilt and running of isocurvature. Thus, CMB-scale isocurvature is suppressed while a characteristic blue-tilted spectrum is generated. As a result, inflationary Hubble scales as large as $H_{\rm inf}\sim 10^{13}\,\mathrm{GeV}$ can be compatible with $f_a\sim 10^{14}$--$10^{16}\,\mathrm{GeV}$, reopening parameter space usually regarded as excluded. We present `observable' benchmarks and a semi-analytic  template that relates the scale-dependence of isocurvature to the geometric lever arm $R/L$, providing a direct phenomenological probe on PQ field-space geometry.
\end{abstract}

\maketitle

\section{Introduction}\label{sec:intro}

The QCD axion provides a dynamical solution to the strong--CP problem and is a well-motivated dark matter (DM) candidate \cite{Peccei:1977hh,Peccei:1977ur,Weinberg:1977ma,Wilczek:1977pj,Marsh:2015xka,DiLuzio:2020wdo,Preskill:1982cy,Abbott:1982af,Dine:1982ah,Badziak:2023fsc}. In the pre-inflationary PQ-breaking scenario, inflation homogenizes the axion field over our observable patch, leaving a coherent background misalignment angle $\theta_i$. However, quantum fluctuations generated during inflation persist as cold-dark-matter isocurvature perturbations, while the axion density also carries the usual adiabatic mode inherited from the primordial curvature perturbation. Current CMB limits on the isocurvature fraction in \cite{Chung:2017uzc,Planck:2018jri,AtacamaCosmologyTelescope:2025nti} therefore translate into a stringent constraint on high-scale inflation as a function of $(f_a,\theta_i)$, sharply disfavoring a vast regime of large axion decay constant $f_a$ (e.g.\ near the GUT scale) together with $H_{\rm inf}\sim 10^{13}\,{\rm GeV}$ \cite{Hertzberg:2008wr}.

The axion isocurvature bound has motivated three main classes of proposals. (i) \emph{Heavy axion during inflation}: suppress fluctuations by making the angular mode heavy, $m_a\gtrsim H_{\rm inf}$, so that superhorizon perturbations are exponentially damped. This can come from a temporarily modified axion potential (enhanced QCD during inflation \cite{Higaki:2014ooa,Choi:2015zra,Co:2018phi,Heurtier:2021rko}, curvature/higher-dimensional terms \cite{Berbig:2024ufe}, monodromy/BF-type masses \cite{Chakraborty:2025lyp}) or SUSY/multi-field transitions that leave a light QCD axion only later \cite{Kearney:2016vqw}. Such mechanisms often require explicit PQ-breaking during inflation and satisfying axion-quality/strong--CP alignment at late times \cite{Kamionkowski:1992mf,Barr:1992qq}. (ii) \emph{Large effective decay constant (non-dynamical)}: suppress fluctuations by displacing the radial field ($R$) associated with the PQ complex scalar, such that $R \gg f_a$ during inflation, and hence the spectrum is suppressed by $(f_a/R)^2$ \cite{Kearney:2016vqw,Kasuya:1996ns,Kasuya:1997td,Kawasaki:2013iha,Chun:2014xva,Nakayama:2015pba,Harigaya:2015hha,Kobayashi:2016qld,Allali:2022yvx,Fairbairn:2014zta}. In minimal quartic PQ models, the radial field is usually assumed to be frozen during inflation with a large displacement and hence the resulting suppressed axion isocurvature spectrum is nearly scale-invariant. This mechanism tends to force tiny quartic coefficient $\lambda$, large excursions (EFT sensitivity), and complicated post-inflationary dynamics (resonance/fragmentation), with the highest-$H_{\rm inf}$ regime especially delicate \cite{Graham:2025iwx}. (iii) \emph{Blue-tilted isocurvature (dynamical)}: use out-of-equilibrium PQ dynamics with a dynamical radial background field, so that the rolling (or rotating) radial background induces a large time-dependent mass for angular fluctuations, suppressing CMB modes while enhancing smaller scales \cite{Kasuya:2009up,Chung:2015pga}. Because the radial field evolves from a large vev towards $f_a$, the suppression weakens over time, producing a blue-tilted isocurvature spectrum. Achieving large blue-tilted spectra over a long range of scales are easiest with effective quadratic radial potential (e.g. Hubble masses in \cite{Kasuya:2009up}). However, pure quartic potential typically yield only a short range unless additional dynamics (e.g. sustained rotation) is present \cite{Chung:2024ctx}. Since the radial field is already driven close to $f_a$ by the end of inflation, this dynamical mechanism can avoid the severe post-inflationary complications associated with a non-dynamical large-displacement scenario.

In this work we present a $U(1)$ conserving alternative model based on field-space geometry. Curved internal field-space metrics are well known to generate effective masses and characteristic scale-dependence for entropic/isocurvature modes in covariant sigma-model systems \cite{Brown:2017osf,Mizuno:2017idt,Chen:2022ccf,Iacconi:2023slv,DeAngelis:2023fdu,Lee:2024bij,DiMarco:2005nq,Renaux-Petel:2014htw}. In this work, we apply this mechanism to the QCD axion, modeled by a spectator PQ complex scalar field with a quartic potential and a spontaneously broken $U(1)_{\rm PQ}$. A hyperbolic field-space metric then exponentially suppresses the observable angular fluctuation without explicit PQ breaking, while the evolving axion mass produces a distinctive blue-tilted isocurvature spectrum on smaller scales. The effective radial mass and the geometric lever arm control the `detectability' of these isocurvatures. As a result, the mechanism reopens viable parameter space for large axion decay constants, e.g. $f_a\lesssim 10^{16}\,{\rm GeV}$, across inflationary scales up to $H_{\rm inf}\sim 10^{13}\,{\rm GeV}$, without requiring the extreme displacements $R_{\rm inf}/f_a\sim 10^{4}$\text{--}$10^{5}$ or tiny quartic couplings $\lesssim \mathcal{O}(10^{-16})$ characteristic of non-dynamical large-displacement scenarios. The upper end of this range, $H_{\rm inf}\sim 10^{13}\,{\rm GeV}$, is especially interesting because it is the regime most relevant for potentially observable primordial gravitational waves. We further provide a compact phenomenological template relating the isocurvature spectral index and running to the field-space curvature scale, thereby promoting measurements of isocurvature scale-dependence to a direct probe of PQ field-space geometry.

The paper is organized as follows. Section~\ref{sec:isocurvature-motivation} reviews the axion isocurvature constraint, Sec.~\ref{sec:hyperbolic} introduces the hyperbolic PQ field-space framework, and Sec.~\ref{sec:Numerical-examples} presents benchmarks and resulting spectra. In Sec.~\ref{sec:post_inflationary_dynamics} we briefly discuss post-inflationary dynamics of the radial field and possible effects on axion isocurvature. We conclude in
Sec.~\ref{sec:conclusion}. Appendix~\ref{app:theta_conservation} discusses conservation of superhorizon isocurvature modes from the end of inflation until QCD epoch, while Appendix.~\ref{sec:setup} gives theoretical derivation for the radial and angular field in a generic metric-space. Appendices~\ref{App:Hyperbolic_UV_example} and \ref{app:template} give a UV-motivated example for the hyperbolic metric, and details of a semi-analytic data-facing template respectively.
%====================================================
\section{The standard QCD axion isocurvature problem
\label{sec:isocurvature-motivation}}

In evaluating the late-time QCD axion isocurvature spectrum, we assume that after the spectator PQ radial mode has relaxed to its vacuum value $f_a$, the superhorizon angular perturbation is conserved until the QCD potential turns on. In Sec.~\ref{sec:post_inflationary_dynamics}, we show that over much of the relevant parameter space in this model, the radial-field evolution between the end of inflation and relaxation to $f_a$ does not modify the conserved far-superhorizon angular fluctuations.

In the pre-inflationary PQ scenario where the PQ complex scalar is a spectator field, if the axion field ($a(x)=f_{\rm eff}\theta(x)$) is light during inflation, its
superhorizon fluctuations are set by the de Sitter result \cite{Baumann:2018muz}
\begin{equation}
\delta a \simeq \frac{H_{\rm inf}}{2\pi},
\qquad
\delta\theta \simeq \frac{\delta a}{f_{\rm eff}(t_\ast)}\,,
\end{equation}
where $f_{\rm eff}(t_\ast)$ is the effective decay constant at CMB horizon exit
(typically $f_{\rm eff}=f_a$ in minimal models), and $\theta_i$ is the misalignment angle. See \cite{Marsh:2015xka} for a review of the discussion in this section.

In the harmonic regime $|\theta_i|\lesssim 1$, the axion density scales as
$\rho_a\propto \theta_i^2$, so the late-time axion isocurvature mode is
\begin{equation}
\mathcal{S}_a \simeq \frac{\delta\rho_a}{\rho_a} \simeq 2\,\frac{\delta\theta}{\theta_i},
\end{equation}where $\delta\theta$ is taken to be the conserved superhorizon angular fluctuations generated during inflation.
If axions constitute a fraction
$r_a\equiv \Omega_a/\Omega_{\rm DM}$ of the total DM and contribute dominantly to the isocurvature signal, then the DM isocurvature power is
\begin{equation}
\Delta_{\mathcal{S},\rm DM}^2(k)\simeq  4r_a^2\frac{\Delta_\theta^2(k)}{\theta_i^2}
\simeq
4r_a^2\frac{H_{\rm inf}^2}{(2\pi)^2 f_{\rm eff}^2(t_\ast)\,\theta_i^2},
\label{eq:PS-generic}
\end{equation}
which is approximately scale invariant in the light-axion limit.
CMB data constrain the isocurvature fraction at the percent level near the pivot
$k_\ast=0.002\,{\rm Mpc}^{-1}$, implying (for $r_a\simeq 1$ and $f_{\rm eff}=f_a$)
\begin{equation}
\frac{H_{\rm inf}}{f_a\,\theta_i} \lesssim 2.8\times 10^{-5}\,.
\label{eq:iso_bound}
\end{equation}

For QCD axions making all of DM via misalignment in a standard thermal history,
$\theta_i$ is fixed by $f_a$ through the relic abundance. Using the usual
scaling $\Omega_a\propto f_a^{7/6}\theta_i^2$ (up to $\mathcal{O}(1)$ uncertainties),
$\Omega_a=\Omega_{\rm DM}$ gives (\cite{DiLuzio:2020wdo,GrillidiCortona:2015jxo})
\begin{equation}
\theta_i\equiv \theta_{i,{\rm QCD}}
\simeq \left(\frac{7.4\times 10^{11}\,{\rm GeV}}{f_a}\right)^{7/12}.
\label{eq:DM_bound}
\end{equation}
This well-known relation assumes the PQ radial field has relaxed to its late-time vacuum
well before the QCD epoch, so the decay constant at the onset of oscillations
is $f_a$.

\begin{figure}[t]
    \includegraphics[scale=0.6]{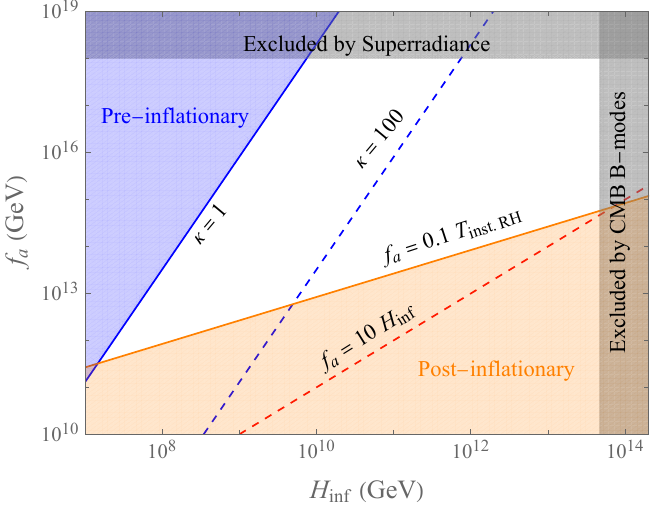}
    \caption{Illustrative constraints in the $(H_{\rm inf},f_a)$ plane for
    pre-inflationary QCD axion DM from misalignment. For the standard flat-metric
    case $f_{\rm eff}=f_a$, the isocurvature bound restricts QCD axion to the blue shaded region.
    Gray bands indicate additional excluded regions. \emph{We focus on how the white
    region can be reopened using geometric suppression.} Our benchmark cases studied in Sec.~\ref{sec:Numerical-examples} are specifically chosen near the most promising value of $H_{\rm inf}\sim10^{13}~\rm{GeV}$. (Redrawn with inspiration
    from \cite{Graham:2025iwx}.)}
    \label{fig:Hinf-fa}
\end{figure}
Fig.~\ref{fig:Hinf-fa} displays the resulting tension in the
$(H_{\rm inf},f_a)$ plane for the usual QCD axion. If the inflationary effective decay constant is
enhanced to $f_{\rm eff}=\kappa f_a$, the bound is relaxed accordingly; combining
Eqs.~\eqref{eq:iso_bound} and \eqref{eq:DM_bound} yields the schematic scaling
\begin{equation}
\frac{f_a}{2.04\times10^{18}{\rm GeV}}
>
\left(\frac{H_{\rm inf}}{10^{10}{\rm GeV}}\frac{1}{\kappa}\right)^{12/5}.
\end{equation} $T_{\rm inst. RH}$ is the instantaneous reheating temperature and the orange solid line marks the upper boundary of the region for PQ restoration post-inflation.
In flat-metric PQ models, realizing simultaneously
$H_{\rm inf}\simeq 10^{13}\,{\rm GeV}$ and $f_a\sim 10^{16}\,{\rm GeV}$ typically
forces $\kappa\gtrsim 10^4$, often into a regime where the spectator/EFT description is difficult to control.

% ============================================================
\section{Hyperbolic PQ field-space and geometric suppression}
\label{sec:hyperbolic}

During inflation we take a quasi-de Sitter background with spatially flat FRW
metric
\begin{equation}
ds^{2}=-dt^{2}+a(t)^{2}d\bm{x}^{2},
\qquad
a(t)\simeq e^{Ht},
\label{eq:frw_metric}
\end{equation}
where $H\simeq H_{\rm inf}$ is approximately constant.
We consider a spectator PQ complex scalar $\Phi$ with a $U(1)$-symmetric
potential and allow a nontrivial internal field-space metric. Adopting a diagonal sigma-model form in radial and angular fields
\begin{equation}
d\sigma^2 = dR^2 + f^2(R)\,d\theta^2,
\end{equation}
so that the action can be written as
\begin{equation}
S=\int d^4x\sqrt{-g}\left[-\frac12(\partial R)^2-\frac12 f^2(R)(\partial\theta)^2 - V(R)\right],
\label{eq:S}
\end{equation}
with $f^2(R)>0$ and no explicit $\theta$-dependence in $V$. The usual flat-metric is readily obtained by taking $f(R)=R$.

Details and derivations in this section are standard and thus deferred to App.~\ref{sec:setup}. We work in the standard non-rotating pre-inflationary PQ background relevant for axion misalignment. This choice  isolates the effect of radial evolution and field-space curvature on axion isocurvature, without introducing the additional dynamics of a rotating PQ condensate. Thus for a vanishing-charge background $\dot\theta_0=0$, the
homogeneous evolution is purely radial and given as
\begin{equation}
\ddot R_0+3H\dot R_0+V_{,R}(R_0)=0,
\end{equation}
and is independent of $f(R)$. The effect of the curved metric appears in the fluctuations. The radial perturbation obeys
\begin{equation}
\ddot{\delta R}+3H\dot{\delta R}-\frac{\nabla^2}{a^2}\delta R
+V_{,RR}(R_0)\,\delta R=0.
\end{equation}
Defining the canonically normalized angular fluctuation $\delta\psi \equiv f(R_0)\,\delta\theta$, one finds
\begin{equation}
\ddot{\delta\psi}+3H\dot{\delta\psi}-\frac{\nabla^2}{a^2}\delta\psi
+m_\psi^2(t)\,\delta\psi=0,
\end{equation}
with the time-dependent effective mass
\begin{equation}
m_\psi^2(t)=\frac{f_{,R}}{f}\,V_{,R}(R_0)-\frac{f_{,RR}}{f}\,\dot R_0^{\,2}.
\label{eq:mpsi_general}
\end{equation}
In flat field-space ($f(R)=R$) one recovers $m_\psi^2 = V_{,R}/R_0$, so the axial fluctuation  has a mass of the same order as the radial mode, and the two evolve with comparable effective masses.

\paragraph*{Hyperbolic metric.}
We now consider a constant negative-curvature field-space,
\begin{eqnarray}
d\sigma^2&=&dR^2+L^2\sinh^2\!\left(\frac{R}{L}\right)d\theta^2,\quad\mbox{where}
\\
f(R)&=&L\sinh\!\left(\frac{R}{L}\right),
\label{eq:hyperbolic_metric}
\end{eqnarray}with curvature scale $L>0$. See Appendix \ref{App:Hyperbolic_UV_example} for a  UV-motivated example that realizes this metric. A canonical example (familiar from
$\alpha$-attractor models \cite{Carrasco:2015rva,Carrasco:2015uma}) is obtained in $N=1$ SUGRA from a ``disk'' K\"ahler potential. The axial mass-squared quantity is then given as
\begin{equation}
m_\psi^2(t)=\frac{1}{L}\coth\!\left(\frac{R_0}{L}\right)V_{,R}(R_0)
-\frac{1}{L^2}\dot R_0^{\,2}.
\label{eq:mpsi_hyp}
\end{equation}
For $R_0\gtrsim L$ and slow-roll background radial evolution, it is useful to write
\begin{equation}
m_\psi^2(t)\approx H_{\rm inf}^2\left(\delta-\frac{\delta^2}{9}\right),
\label{eq:mpsi_hyp_2}
\end{equation}
where
\begin{equation}
\delta\equiv \xi\,\frac{m_R^2}{H_{\rm inf}^2},
\qquad
\xi\equiv \frac{R_0}{L},
\qquad
m_R^2\equiv \frac{V_{,R}(R_0)}{R_0}.
\label{eq:xi_defn}
\end{equation}
The leading geometric contribution is enhanced by the curvature lever arm
$\xi=R_0/L$, while the subleading negative term encodes the reduction
of transverse masses from negative curvature when $\dot R_0\neq 0$.
Unlike for a flat-metric, the axial mode can become $\mathcal{O}(H)$-heavy even during radial slow-roll. The time dependence generically
produces blue-tilted isocurvature: modes exiting earlier (larger scales)
are most suppressed, while later-exiting modes (smaller scales) are less
suppressed. From Eq.~(\ref{eq:mpsi_hyp}), we also note that large radial velocities (underdamped motion) tend to destabilize the angular mode.

% ======================================================
\section{Benchmark models and isocurvature spectrum}
\label{sec:Numerical-examples}

We adopt a quartic PQ potential
\begin{equation}
V(R)=\frac{\lambda}{4}\left(R^{2}-f_{\rm PQ}^{2}\right)^{2},
\qquad
f_{\rm PQ}\simeq f_a,
\label{eq:V_mexicanhat}
\end{equation}
with hyperbolic metric Eq.~(\ref{eq:hyperbolic_metric}) and work in the vanishing charge sector $\dot\theta_0=0$. The quartic potential is a  standard minimal renormalizable potential used in PQ models \cite{Marsh:2015xka}. Our purpose is precisely to ask how the familiar pre-inflationary PQ scenario is modified when the PQ field space has nontrivial curvature. In this sense, the quartic model with a zero charge should be viewed as a minimal deformation of the conventional quartic PQ setup, rather than as a specialized construction.

We evolve the background spectator radial and axial fields during inflation using $N\equiv \ln a$. For each comoving mode $k$, we solve the full linearized fluctuation equations numerically, starting from Bunch-Davies initial conditions deep inside the horizon, $k/(aH)\gg 1$, and continuing the evolution through horizon crossing and into the superhorizon regime. The late-time angular power spectrum at the end of inflation is
\begin{equation}
\Delta_{\theta}^{2}(k)\equiv\frac{k^{3}}{2\pi^{2}}\,|\delta\theta_{k}|^{2},
\qquad
\delta\theta_{k}=\left.\frac{\delta\psi_{k}}{f(R_{0})}\right|_{\rm late-time}.
\label{eq:theta_power_def}
\end{equation}
Based on the assumptions detailed at the beginning of Sec.~\ref{sec:isocurvature-motivation}, in the harmonic regime the total DM isocurvature spectrum is given by Eq.~(\ref{eq:PS-generic}). For QCD axion DM from
misalignment, we eliminate $\theta_i$ using Eq.~(\ref{eq:DM_bound}).

\subsection{Detectability of isocurvature signal}
Eq.~\ref{eq:theta_power_def} shows that in the large-curvature regime, $R/L \gg 1$, the hyperbolic metric gives \[\Delta_{\theta}^{2} \sim \left(H_{\rm inf}/L\right)^{2} e^{-2R_0/L}.\] Thus, once $R/L \gg 10$, the angular fluctuations are exponentially suppressed and the resulting isocurvature signal is phenomenologically negligible. For representative values $H_{\rm inf}\lesssim L$, near-term observability of the isocurvature therefore requires the relevant field range to satisfy $R\lesssim \mathcal{O}(10)L$.

Within this potentially detectable window, the shape of the isocurvature spectrum is controlled primarily by the time evolution of the radial background. There are two qualitatively different regimes.

First, if the radial mode is strongly overdamped,
\[m_R^2 \equiv \frac{V_{,R}}{R} \lesssim \mathcal{O}(0.01)H_{\rm inf}^2,\]
then the radial background changes very little over the observable range of inflationary e-folds. In this limit, the effective normalization of the angular fluctuation is nearly time independent. Combining Eqs.~(\ref{eq:mpsi_hyp_2}) and (\ref{eq:xi_defn}), for $R/L\lesssim \mathcal{O}(10)$, the geometry-induced angular mass remains small,
\[m_\psi^2 \lesssim \mathcal{O}(0.1)H_{\rm inf}^2,\]
and the resulting isocurvature spectrum is approximately scale invariant, with negligible running. Such a signal can still be visible on CMB scales in future experiments, but because the spectrum is suppressed compared to adiabatic fluctuations and has little scale dependence, it does not produce a distinctive enhancement on smaller scales.

Second, if the radial field undergoes a mild slow-roll evolution,
\[m_R^2 \lesssim \mathcal{O}(0.1)H_{\rm inf}^2,\]
the radial background can change by an order-one amount over the observable e-folds. The angular sector then experiences a time-dependent geometry-induced mass, typically of order
$m_\psi^2 \sim H_{\rm inf}^2.$
This time dependence suppresses the modes that exit earliest, corresponding to CMB scales, while allowing larger power in modes that exit later. The resulting axion isocurvature spectrum is therefore significantly blue tilted: it can evade CMB-scale bounds while becoming enhanced on smaller scales. Because the radial background evolves continuously during horizon exit, the tilt itself is also scale dependent, leading to a measurable running. This running provides a characteristic signature of the geometric mechanism, distinguishing it from a simple scale-invariant suppression of axion fluctuations.

This is qualitatively different from mechanisms that simply suppress axion
fluctuations by a nearly scale-independent factor. The hyperbolic mechanism does
not merely evade isocurvature constraints by hiding the signal; it opens a viable
window in which the CMB-scale isocurvature is geometrically suppressed, while the
smaller-scale spectrum remains large enough to be potentially observable. This
correlated CMB-scale suppression and growth toward smaller scales is the central
phenomenological signature of the model. Moreover, the characteristic running
predicted by the geometric template provides an additional discriminator from
other blue-tilted isocurvature scenarios.

Therefore, we choose the benchmark points to lie near $R/L\sim 10$. In this regime the exponential suppression from the hyperbolic metric is strong enough that $\Delta_{\mathcal S}^2$ on CMB scales remains safely below current bounds, while the subsequent radial evolution can still generate an enhancement on smaller scales, with the power scaling as $k^{\sim1}$.

\begin{table}[h]
\centering
\setlength{\tabcolsep}{5pt}
\renewcommand{\arraystretch}{1.15}
\begin{tabular}{lccccc}
\hline
 & $\lambda$ & $f_a/H_{\rm inf}$ & $R_i/H_{\rm inf}$ & $N_{\rm inf}$ & $L/H_{\rm inf}$ \\
\hline
$\textsf{B1}$ & $3.85\times10^{-8}$ & $10^{1}$ & $3.3\times10^{3}$ & $55$ & $2.35\times10^{2}$ \\
$\textsf{B2}$ & $7.70\times10^{-7}$ & $10^{1}$ & $7.7\times10^{2}$ & $55$ & $4.70\times10^{1}$ \\
$\textsf{B3}$ & $3.85\times10^{-10}$ & $10^{3}$ & $3.3\times10^{4}$ & $55$ & $2.35\times10^{3}$ \\
\hline
\end{tabular}
\caption{Benchmark parameter choices.}
\label{tab:benchmarks}
\end{table}

We consider three benchmark points $\textsf{B1}$-$\textsf{B3}$
(Tab.~\ref{tab:benchmarks}), chosen such that the initial curvature lever arm
$\xi_i\equiv R_i/L$ is $\sim\mathcal{O}(10)$. For definiteness, and to illustrate the most observationally interesting and most constrained regime, we evaluate the benchmarks at a high inflationary scale, $H_{\rm inf}\sim10^{13}\,{\rm GeV}$, relevant for potentially detectable primordial gravitational waves. The mechanism is not restricted to this value of $H_{\rm inf}$; lower inflationary scales are also viable and generally face weaker CMB isocurvature constraints, thus typically requiring a smaller curvature lever arm $R/L$.

Furthermore, the benchmark points are selected not only to demonstrate viable spectra at high $H_{\rm inf}$, but also to highlight two useful structural features of the parameter space. The near-degeneracy of \textsf{B1} and
\textsf{B3} is intentional. When expressed in units of $H_{\rm inf}$, the spectator axial isocurvature power has
an approximate rescaling symmetry,
\begin{equation}
R\rightarrow \frac{R}{x},\qquad
L\rightarrow \frac{L}{x},\qquad
f_{a}\rightarrow \frac{f_{a}}{x^{24/14}},\qquad
\lambda\rightarrow \lambda\,x^{2},
\label{eq:iso_scaling_map}
\end{equation}
which leaves the dimensionless geometry parameter ($R/L$) and the radial mass in Hubble
units approximately invariant, yielding nearly identical spectra for \textsf{B1}
and \textsf{B3} despite different values of $f_a$.  Benchmark \textsf{B2}, by contrast, is chosen to illustrate that viable spectra are not confined to this nearly degenerate scaling trajectory. At fixed $f_a/H_{\rm inf}=10$, it varies the combination ($\lambda,~L/H_{\rm inf}$), thereby showing the broader range of parameter choices capable of producing suppressed CMB-scale isocurvature together with enhanced small-scale power.

\subsection{Spectator condition and stochastic drift}
The validity of the spectator approximation relies on the PQ sector remaining subdominant to the inflaton-driven expansion. We verify this by calculating the energy density ratio:
\begin{equation}\frac{\rho_{R}}{\rho_{\rm inf}} \simeq \frac{\lambda R_i^4}{12 H_{\rm inf}^2 M_p^2} \ll 1.
\end{equation}For the upper bound of the inflationary scale, $H_{\rm inf} \simeq 10^{13}\,\text{GeV}$, our benchmarks $B_1,~B_2,$ and $B_3$ yield $\rho_R/\rho_{\rm inf} \sim 10^{-7}\text{--}10^{-3}$. Because $\rho_R$ remains a subdominant fraction of the total energy budget, its contribution to the background geometry and consequently the generation of metric perturbations is negligible. Furthermore, as we have checked and also show in Sec.~\ref{sec:results}, the linear perturbation $\delta \theta_k$ remains within the perturbative regime, $\mathcal{P}_{\delta \theta}^{1/2} \ll 1$, for all relevant scales. The backreaction of these fluctuations on the background geometry is proportional to $(\delta \theta)^2$, which is significantly smaller than the background energy density. Thus, the PQ sector remains a passive spectator, and the coupling to curvature perturbations through metric backreaction remains below the threshold of phenomenological importance.

We further confirm the validity of our classical trajectory treatment that deterministic drift dominates over quantum diffusion. For a field evolving in a de Sitter background, the classical displacement over one Hubble time, $\Delta R_{\text{cl}} \simeq |V_{,R}|/(3H^2)$, exceeds the quantum diffusion step, $\Delta R_{\text{stoch}} \simeq H/(2\pi)$, provided $|V_{,R}|/H^3 \gtrsim \mathcal{O}(1)$. Given our benchmark parameters $m_R^2/H^2 \gtrsim 0.01$ and $R \gtrsim 10\text{--}100 f_a$ with $f_a \geq 10H_{\rm inf}$, the classical slope $V_{,R} = m_R^2 R$ satisfies $|V_{,R}|/H^3 \sim (m_R^2/H^2)(R/H) \gtrsim \mathcal{O}(1)$. Consequently, the classical drift velocity significantly exceeds the stochastic diffusion rate, rendering the field trajectory essentially deterministic. Furthermore, the relative diffusion $\Delta R_{\text{stoch}}/R \lesssim \mathcal{O}(10^{-2})$ is a sub-percent effect, ensuring that the field's evolution is not smeared by random fluctuations and that our classical benchmark trajectories provide a robust description of the predicted isocurvature spectrum and its scale-dependent tilt.

\subsection{Results}\label{sec:results}

\begin{figure}[h]
\includegraphics[scale=0.5]{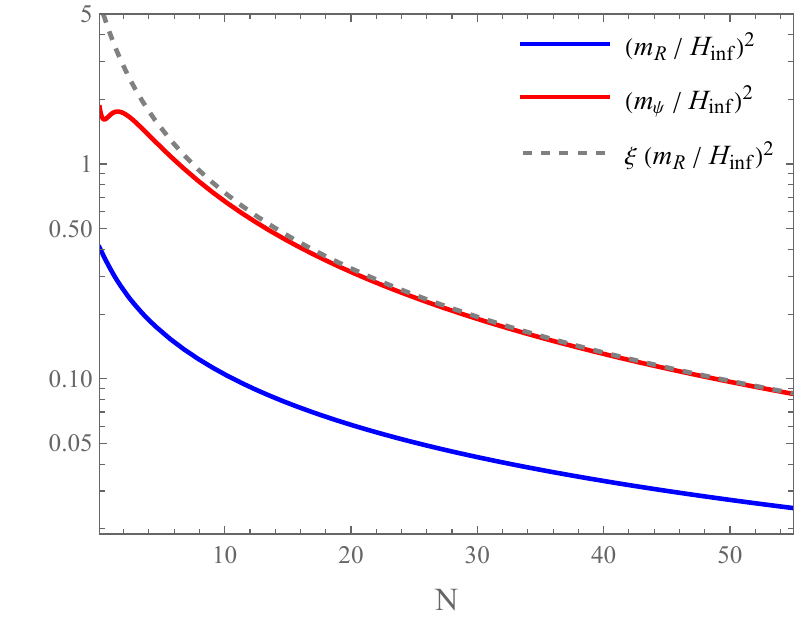}
\caption{Time-dependent mass-squared of the
 angular fluctuation ($m_\psi^2$) and background radial mass-scale  ($m_R^2$) for \textsf{B1}.
\label{fig:mass_sq}}
\end{figure}
Fig.~\ref{fig:mass_sq} shows the mass history for \textsf{B1}.
As the radial field slow-rolls toward its vacuum, while the hyperbolic metric induces a much larger geometric axial mass.
Initially $m^2_R\sim \mathcal{O}(0.1)H_{\rm inf}^2$ while $m_\psi^2 \sim \mathcal{O}(H_{\rm inf}^2)$, so axial modes exiting early are
damped. As $R_0$ decreases and $\dot R_0$ redshifts, $m_\psi^2/H_{\rm inf}^2$ falls and the damping weakens for later-exiting modes. In the regime $\delta=\xi(m_R/H_{\rm inf})^2\lesssim 1$, the dominant behavior is $m_\psi^2\simeq \xi\,m_R^2$, i.e.\ the angular mode is enhanced relative to the flat case by the curvature lever arm $\xi=R/L$.
\begin{figure}[t]
\includegraphics[scale=0.5]{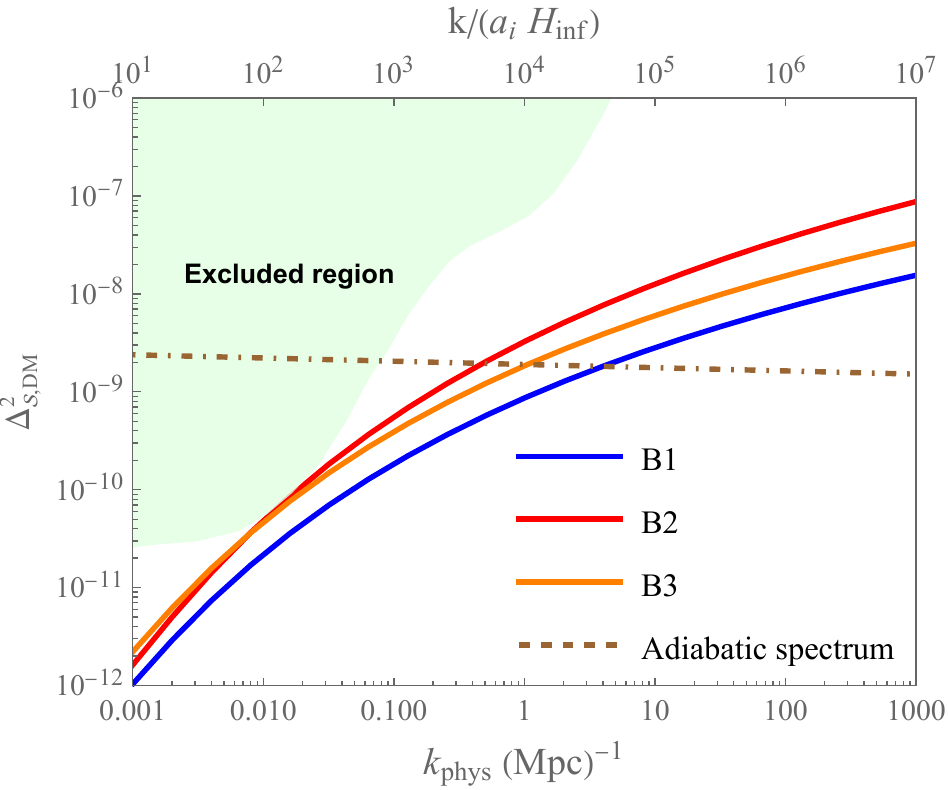}
\caption{Total DM isocurvature spectra for the benchmarks, assuming QCD axions
constitute all DM via misalignment ($r_a=1$). We take $H_{\rm inf}=10^{13}\,{\rm GeV}$. Adiabatic spectrum plotted for $A_s=2.1\times10^{-9}$ and $n_s=0.967$ \cite{Planck:2018jri}. For comparison, we also show the current observational bounds from \cite{Co:2026klr} on scales $k\leq 10~\rm{Mpc}^{-1}$. Bounds on even smaller-scales are generally much weaker than at $k = 10~\rm{Mpc}^{-1}$.}
\label{fig:isocurvature_v1}
\end{figure}

Fig.~\ref{fig:isocurvature_v1} shows the resulting conserved superhorizon total DM isocurvtaure spectrum $\Delta_\mathcal{S}^2(k)$ for the three benchmarks cases.
We take $k/(a_iH_{\rm inf})=10$ to correspond to the CMB pivot scale
$k_\ast\simeq 0.001,{\rm Mpc}^{-1}$. Power is strongly suppressed on the
largest CMB scales because those modes exit while $f(R)$ is exponentially enhanced, so that the observable fluctuation $\delta\theta=\delta\psi/f(R)$ is strongly suppressed at horizon exit \cite{Lee:2024bij}. As $R$ slow-rolls during the radial evolution, $f(R(t_k))$ decreases and
smaller scales are progressively less suppressed. The outcome is a blue-tilted
isocurvature spectrum with a scale-dependent running over the range of modes that probe the
evolution of $R_0$, approaching a quasi-plateau once the
angular mode becomes light. Equivalently and rather more accurately, this behavior can be described in terms of the time-dependent effective mass $m_\psi^2(t)$ during radial slow-roll, which governs evolution of $\delta\psi$ while leaving the ratio $\delta\psi/f(R)=\delta\theta$ nearly conserved in the slow-roll regime outside the horizon. For comparison, we also show the current observational bounds on the total DM isocurvature spectrum for a running spectrum from \cite{Co:2026klr}, which is based on combined CMB and UV luminosity function datasets. In particular, the benchmarks studied here yield an isocurvature-to-adiabatic ratio of order unity near $k\sim 0.5\,{\rm Mpc}^{-1}$, consistent with recent CDM-isocurvature constraints in \cite{Co:2026klr}. For more strongly blue-tilted spectra within our framework, an order-unity ratio can be achieved already near $k \sim 0.1\,\mathrm{Mpc}^{-1}$, while remaining consistent with current observational bounds.

The above results highlight that hyperbolic geometry can reopen parameter space that is excluded in the flat-metric case. In our benchmarks, we focussed on $H_{\rm inf}=10^{13}\,{\rm GeV}$ and $f_a \sim 10^{14}\text{--}10^{16}\,{\rm GeV}$,
as it lies near current upper limits and is the most constrained
regime for isocurvature. For smaller $H_{\rm inf}$ at a fixed $f_a$, the axion fluctuation amplitude is relatively lower, so CMB isocurvature constraints are relatively weaker and thus viable parameter space generally opens up with a smaller required curvature lever arm $R/L$. In contrast, for $f(R)=R$, satisfying the CMB bound at $H_{\rm inf}=10^{13}\,{\rm GeV}$ typically requires extreme radial excursion ($R_i/f_a \sim 10^{4}\text{--}10^{5}$), implying super-Planckian displacements.

\begin{figure}[t]
\includegraphics[scale=0.55]{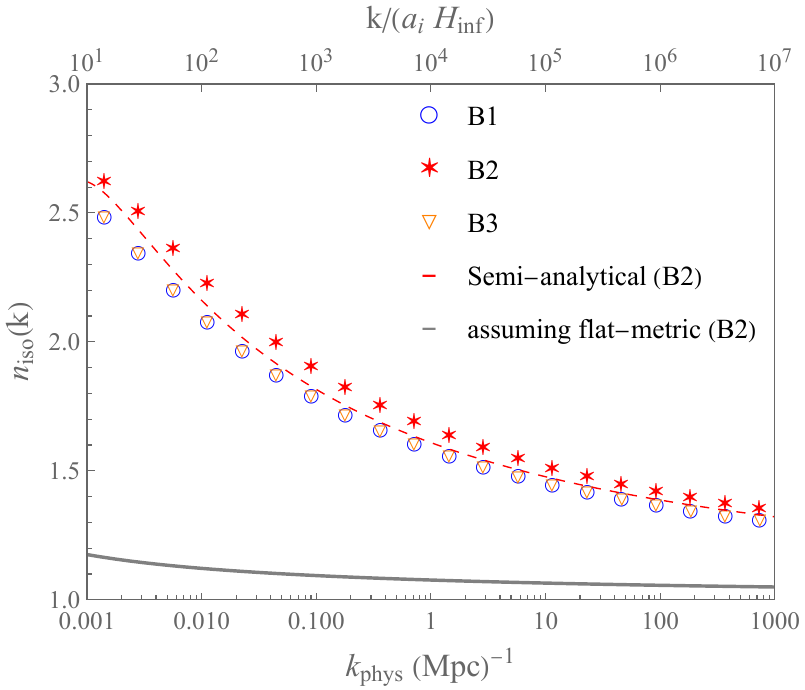}
\caption{Isocurvature spectral index $n_{\rm iso}(k)$ inferred from the numerical
spectra. The red dashed curve shows the semi-analytic estimate in Eq.~(\ref{eq:niso}) for \textsf{B2}. The gray curve indicates the corresponding flat-metric expectation.}
\label{Fig:spectral_index}
\end{figure}
% =======================
% Main text (template + highlights)
% =======================

\subsection{Running power spectrum template}\label{sec:template}
Fig.~\ref{Fig:spectral_index} shows the isocurvature tilt and its
running inferred from numerical spectra shown in Fig.~\ref{fig:isocurvature_v1}.  For a standard running spectrum, we write the power-law parameterization as
\begin{equation}
\Delta^2_{\mathcal{S}} = A_{\rm iso} \left(\frac{k}{k_p}\right)^{n_{\rm iso}(k) - 1},\label{eq:running_spectrum}
\end{equation}where $k_p$ is the pivot scale usually set at $0.05\,\rm{Mpc}^{-1}$. When $m^2_\psi/H^2_{\rm inf}$ varies slowly around horizon exit, the tilt
is well approximated by the massive-mode estimate
\begin{equation}
n_{\rm iso}(k)\simeq 4-2\sqrt{\frac{9}{4}- \mu(t(k))},
\label{eq:niso}
\end{equation}where we introduce \[ \mu(t) = m_\psi^2(t)/H_{\rm inf}^2\]
with $t(k)$ fixed by $k=a(t)H_{\rm inf}$. Evaluating $m_\psi^{2}(t)$ by numerically solving the background radial field and substituting into Eq.~(\ref{eq:niso}) reproduces the approximate $n_{\rm iso}(k)$ as shown by the red dashed curve in Fig.~\ref{Fig:spectral_index} for \textsf{B2}. For clarity, we only show the result for \textsf{B2}.

In the slow-roll geometric domain, the hyperbolic metric ties the axial mass to
the radial background through $\xi\equiv R/L$. For quartic PQ slow-roll, we give the following scale-dependent mass template (see App.~\ref{app:template} for complete derivation)
\begin{equation}
\mu(k)\simeq
\mu_p
\left[
1+\frac{2}{3}\left(\frac{\mu_p}{\xi_p}\right)
\ln\!\left(\frac{k}{k_p}\right)\right]^{-3/2}
\label{eq:mpsik_template}
\end{equation}$\mu_p=\mu(k_p)$ and $\xi_p=\xi(k_p)$. From the running spectrum defined in Eq.~\ref{eq:running_spectrum}, we infer that
\begin{eqnarray}
n_{\rm iso}(k_p) &=& 4 -  \sqrt{9 - 4\mu_p}, \\
\alpha_{\rm iso}(k_p)=\left.\frac{dn_{\rm iso}(k)}{d\ln k}\right|_{k_p} &=&   -\frac{4 \mu_p^2}{\xi_p\sqrt{9-4 \mu_p} }.
\end{eqnarray}Since the tilt $n_{\rm iso}(k_p)$ is uniquely determined by $\mu_p$, the running $\alpha_{\rm iso}(k_p)$ is dictated by $\xi_p$. For $\mu_p \lesssim 1$ and $\xi_p \gg 1$, the running parameter is approximately constant. Hence $\alpha(k) \approx	\alpha(k_p)$ for $k \lesssim k_p \exp(\xi_p/\mu_p)$. Thus, while $\mu_p$ controls the spectral tilt at $k_p$, $\xi_p\equiv R_p/L$ controls the overall shape over a broad range of scales.

The salient point is the functional form. In the hyperbolic regime one has
$m_\psi^2\propto \xi\,m_R^2\propto R\,m_R^2$. The hyperbolic lever arm contributes an addtional factor of $R$ to the canonical axial mass. For a monomial-like slow-roll potential, $V(R)\propto R^q$ with ($q>2$), the radial field follows a tracking solution with $R(N)\propto (N-N_p)^{-1/(q-2)}$. In the flat-metric case, the axial mass-squared quantity is thus independent of the monomial-shape and given by the relation $m^2_\psi(N) \propto (N-N_p)^{-1}$. But in the hyperbolic case, $m^2_\psi(N) \propto (N-N_p)^{-(q-1)/(q-2)}$. Using the relation $\ln k \sim N$, we obtain for quartic potentials
\begin{eqnarray}
m_\psi^2(N)&\propto& (N-N_p)^{-3/2}
\\
m_\psi^2(k)&\propto& \left[\ln\!\left(\frac{k}{k_p}\right)\right]^{-3/2}.
\label{eq:minus3half_highlight}
\end{eqnarray}
This is steeper than the familiar $ (N-N_p)^{-1}$ scaling
that arises for a broad class of slow-roll backgrounds (e.g. monomial potentials) \cite{Lyth:1998xn}. The extra half-power is the geometric fingerprint. 

Fitting the numerical spectral index $n_{\rm iso}(k)$ in  Fig.~\ref{Fig:spectral_index} with our template in  Eqs.~\eqref{eq:niso} and \eqref{eq:mpsik_template} using least chisquare fitting method
gives $\xi_p\simeq 9.26\pm0.17$ for \textsf{B1, B3} and $\xi_p\simeq 10.88\pm0.28$ for \textsf{B2}, and explains the apparent similarity of the isocurvature spectral shapes in Fig.~\ref{fig:isocurvature_v1}. Thus, the scale-dependence is well captured by Eqs.~\eqref{eq:niso} and \eqref{eq:mpsik_template} which define a predictive template in which the running directly measures the field-space curvature through $\xi_p$.

\section{Post-Inflationary Dynamics}
\label{sec:post_inflationary_dynamics}
Following inflation, as the Hubble parameter drops, the radial field becomes dynamical (underdamped) again and must relax toward its global, zero-temperature minimum at $f_a$. 

A primary theoretical concern in such dynamical decay constant models is that rapid post-inflationary oscillations of the radial field can induce explosive parametric resonance (PR) in the axion angular field, $\theta$. If this resonance is sufficiently strong, the energy transferred to the axion fluctuations, $\delta\theta_k$, can backreact on the zero-mode condensate. This leads to complete fragmentation of the field, a non-thermal restoration of the Peccei-Quinn (PQ) symmetry, and the subsequent erasure of the precise isocurvature conditions established during inflation.

However, as recently demonstrated in \cite{Graham:2025iwx}, this catastrophic fragmentation is robustly avoided provided the initial radial displacement is moderate. Specifically, our benchmarks focus on the parameter space where the radial displacement at the end of inflation ($t_{\rm end}$) is bounded by $R(t_{\rm end})/f_a \lesssim 10 - 100$. As previously highlighted, this constitutes a distinct advantage of the geometric mechanism. Unlike flat-metric PQ fields where $R(t_{\rm end})/f_a \gtrsim 10^{4}$ to suppress isocurvature for high-scale inflation and large axion decay-constant, the exponential suppression induced by the hyperbolic geometry ensures that phenomenologically safe isocurvature bounds can be achieved without requiring an exceptionally large radial displacement. 

The post-inflationary field theory in our framework governing the radial oscillations in the standard symmetry-breaking potential, $V(R) = \frac{\lambda}{4}(R^2 - f_a^2)^2$, is identical to that in \cite{Graham:2025iwx}. Therefore, the detailed numerical and analytical results in \cite{Graham:2025iwx} regarding PR are transferable to our scenario. As highlighted in their work, the safety of the $R(t_{\rm end})/f_a \lesssim 100$ regime is rooted in the dynamics of the radial field's relaxation (see Sec.~4.1 and Sec.~4.2 of \cite{Graham:2025iwx}).

For $R \gg f_a$, the radial field's envelope redshifts rapidly with the cosmological scale factor, $R \propto a^{-1}$. For an initial amplitude $R(t_{\rm end}) \sim 100 f_a$, the field reaches the minimum in roughly $\Delta N \approx \ln(100) \sim 4.6$ $e$-folds of expansion. Significant backreaction requires the radial/axial fluctuations to undergo massive exponential amplification (requiring a large Floquet exponent integrated over many oscillations). Because the radial field redshifts to $f_a$ so efficiently, the duration of the high-amplitude quartic regime is insufficient for subhorizon modes to grow to the non-linear regime. Once the radial envelope drops to $f_a$, the field transitions to oscillating around the quadratic minimum. As detailed in Appendix A.1 of \cite{Graham:2025iwx}, this places the system strictly in the narrow-band resonance regime. Hubble expansion  rapidly sweeps the comoving scales out of the narrow instability bands and permanently quenches the resonance.

Consequently, for $R(t_{\rm end})/f_a \lesssim 10\text{--}100$, parametric resonance is weak and transient. The zero-mode condensate is undisturbed by backreaction, precluding both field fragmentation and non-thermal PQ restoration. The superhorizon zero-mode remains coherent throughout the post-inflationary epoch, strictly conserving the inflationary isocurvature predictions. Furthermore, the highly relativistic axions generated by this weak PR represent a negligible fraction of the total energy density, smoothly redshifting as dark radiation without disrupting the standard axion cold dark matter relic abundance. A detailed numerical analysis will be addressed in future work.

\section{Conclusion}\label{sec:conclusion}

We have demonstarted that the intrinsic geometry of the PQ field manifold provides a
 symmetry-preserving solution for the axion isocurvature
problem. The curved (hyperbolic) field-space enhances the effective decay constant and can suppress CMB-scale isocurvature. The angular fluctuation acquires a geometric, time-dependent effective mass during inflation, without any explicit PQ breaking in the axion potential. 

The phenomenology is governed by the geometric lever arm $R/L$ and radial dynamics. For moderate $R/L \lesssim \mathcal{O}(10)$, the geometric mechanism not just evades the axion isocurvature bound but can also yield potentially observable signal on smaller scales. Thus, the signal can remain detectable and CMB-compatible while radial slow-roll enhances small-scale power and induces measurable running. If the radial mode is strongly overdamped, the spectrum is nearly scale-invariant with negligible running, whereas very large $R/L$ suppresses the signal below measurable limits. In our benchmark realizations with $H_{\rm inf}=10^{13}\,{\rm GeV}$, this mechanism reopens parameter space at large $f_a$ that is typically excluded in flat field-space.

Beyond enlarging the viable QCD-axion window at high inflationary scales, our
results highlight that isocurvature scale-dependence can directly probe field-space curvature. We provided a compact template linking the running of the isocurvature tilt $\alpha_{\rm iso}$ to a curvature-controlled parameter $\xi_p$, enabling model-agnostic tests of the mechanism and straightforward incorporation into future analyses. 

Several broad directions are motivated by this perspective. On the theory side, constructing controlled UV realizations of hyperbolic metrics in PQ models, and extending the framework to more general target spaces (e.g. non-constant curvature or non-diagonal metrics) are important. On the dynamics side, nontrivial background motion such as $\dot\theta\neq0$ or post-inflationary evolution may introduce additional observational constraints, or signatures beyond the power spectrum \cite{Renaux-Petel:2021yxh,Firouzjahi:2021lov}. Within rotating-field mechanisms such as Affleck-Dine dynamics in baryogenesis (\cite{AFFLECK1985361}), axiogenesis \cite{Co:2019wyp} and kinetic-misalignment in \cite{Co:2019jts}, hyperbolic geometry can support large angular velocities. Finally, the predicted enhancement of sub-CMB power motivates systematic confrontation with small-scale probes, including Lyman-$\alpha$ \cite{Buckley:2025zgh}, UV luminosity functions \cite{Yoshiura:2020soa}, 21-cm measurements \cite{Minoda:2021vyw}, and strong-lensing substructure \cite{Dalal:2002su,Wu:2024tdh}, offering a
direct observational window onto PQ field-space geometry.

\begin{acknowledgments}
We thank Peter Graham, Davide Racco, Raymond Co, and Tomo Takahashi for helpful comments and correspondence. We also thank the referees for their suggestions which significantly improved the structure of this paper.
\end{acknowledgments}

\appendix

\section{Conservation of the far-superhorizon angular perturbation}
\label{app:theta_conservation}

In this appendix, we show that within the zero-charge and unbroken-PQ regime, homogeneous post-inflationary radial evolution does not modify the far-superhorizon angular perturbation. This justifies using the inflationary phase fluctuation as the precise initial condition for the late-time QCD axion misalignment calculation.

We consider the general PQ action given in Eq.~\ref{eq:S} and assume $V(R)$ is $U(1)_{\rm PQ}$-symmetric and contains no explicit $\theta$-dependence. The exact angular equation of motion is the conservation equation for the Noether current associated with the shift symmetry $\theta\to \theta+{\rm const}$,
$$ \nabla_\mu J^\mu = 0, \qquad J^\mu=f^2(R)\nabla^\mu\theta . $$
On a spatially flat FRW background, this yields
$$ \frac{d}{dt}\left(a^3 f^2(R)\dot\theta\right) - a f^2(R)\nabla^2\theta = 0 . $$
For a homogeneous background, the conserved charge is therefore $Q_\theta \equiv a^3 f^2(R_0)\dot\theta_0$. The pre-inflationary axion misalignment setup used in this work assumes zero initial rotation, meaning $Q_\theta=0$, or equivalently $\dot\theta_0=0$. In this sector, the background evolution is purely radial, while the angular perturbation obeys, to linear order:
$$ \ddot{\delta\theta_k} + \left(3H+2\frac{\dot f}{f}\right)\dot{\delta\theta_k} + \frac{k^2}{a^2}\delta\theta_k = 0 , $$
where $f\equiv f(R_0(t))$. Notice that no assumption of slow-roll is required here. The radial field may slow-roll or oscillate rapidly; this simply modifies the time-dependent friction coefficient $3H+2\dot{f}/f$.

For a far-superhorizon mode, $k/(aH)\ll1$, the gradient term is negligible, yielding
$$ \frac{d}{dt}\left(a^3 f^2(R_0)\dot{\delta\theta_k}\right) \simeq 0 \implies a^3 f^2(R_0)\dot{\delta\theta_k} = C_k , $$
with $C_k$ constant. By the end of inflation, superhorizon modes $\delta\theta_k$ have a velocity $\dot{\delta\theta_k} \propto (k/(aH))^2 \delta\theta_k$ which is exponentially suppressed for CMB-scale modes. Hence, the integration constant $C_k$ is negligible. The phase perturbation relevant for CMB-scale isocurvature is perfectly conserved on far-superhorizon scales:
$$ \delta\theta_k \simeq {\rm constant} \qquad (k\ll aH,\; Q_\theta=0). $$

It is conceptually useful to contrast this conserved phase with the canonically normalized angular fluctuation, $\delta\psi_k = f(R_0)\delta\theta_k$. Because the decay constant $f(R_0)$ is dynamically evolving, the canonical field $\delta\psi_k$ is not trivially conserved. However, the late-time axion misalignment angle is determined by the phase variable $\theta$, not the instantaneous canonical variable $\psi$. Once the radial field relaxes to the vacuum ($R_0 \to f_a$), the physical axion fluctuation becomes $\delta a_k = f_a\delta\theta_k$. Thus, for modes that remain superhorizon until after the radial field has settled, the conserved $\delta\theta_k$ computed during inflation exactly corresponds to the fluctuation entering the late-time QCD axion abundance.

By the separate-universe argument, a CMB-scale mode remains far outside the horizon during the post-inflationary relaxation of the PQ radial field. Each local patch evolves as an independent, homogeneous universe with phase $\theta_0+\delta\theta$. Because there is no $\theta$-dependent force before the QCD potential turns on, the radial evolution is identical across all patches, preserving the relative phase displacement.

Finally, we briefly comment on the severe post-inflationary disruptions that plague models with ultra-large initial displacements. If PR were sufficiently explosive, the radial condensate would fully fragment, leading to a non-thermal restoration of the Peccei-Quinn symmetry. In such a scenario, the angular variable is no longer cleanly inherited from its inflationary value; topological defects (cosmic strings and, subsequently, domain walls) would form, and the system would be driven into the standard post-inflationary axion scenario, irrevocably erasing the precise isocurvature initial conditions.

However, as established in Section \ref{sec:post_inflationary_dynamics}, these pathological outcomes are not bound to occur in our model. If the  final radial displacement before post-inflationary oscillations is confined to $R(t_{\rm end})/f_a \lesssim 10\text{--}100$, PR is strictly weak and transient (\cite{Graham:2025iwx}). The condensate remains robustly intact, and PQ symmetry is not restored. While a full lattice analysis of the highly nonlinear catastrophic fragmentation regime is an interesting subject for future work, it is mostly bypassed in our parameter space. Consequently, local subhorizon dynamics would not disrupt the long-wavelength angular mode, and the inflationary far-superhorizon phase perturbation remains conserved through the post-inflationary epoch.

\section{Theoretical setup}\label{sec:setup}

We start from a canonically normalized complex scalar field $\Phi$ with a
$U(1)$-symmetric potential $V(|\Phi|)$ and no explicit $U(1)$-breaking terms,
\begin{equation}
S=\int d^{4}x\,\sqrt{-g}\left[-g^{\mu\nu}\partial_{\mu}\Phi^{*}\partial_{\nu}\Phi
-V\big(|\Phi|\big)\right],
\label{eq:canonical_complex_action}
\end{equation}
where $g_{\mu\nu}$ is the spacetime metric and $g\equiv\det g_{\mu\nu}$.
It is convenient to parameterize the complex field in polar coordinates,
$(R,\theta)$, and the kinetic term becomes
\begin{equation}
-g^{\mu\nu}\partial_{\mu}\Phi^{*}\partial_{\nu}\Phi
=-\frac{1}{2}g^{\mu\nu}\left(\partial_{\mu}R\,\partial_{\nu}R
+R^{2}\,\partial_{\mu}\theta\,\partial_{\nu}\theta\right),
\end{equation}
such that the action can be written in sigma-model form,
\begin{equation}
S=\int d^{4}x\,\sqrt{-g}\left[-\frac{1}{2}G_{IJ}(\phi)\,g^{\mu\nu}
\partial_{\mu}\phi^{I}\partial_{\nu}\phi^{J}-V(R)\right],
\label{eq:sigma_model_action}
\end{equation}
where $\phi^{I}=(R,\theta)$ and the internal metric defines the field-space line
element
\begin{align}
d\sigma^{2}
&=G_{IJ}(\phi)\,d\phi^{I}d\phi^{J}
\label{eq:canonical_field_metric}\\
&=G_{RR}\,dR^{2}+G_{\theta\theta}\,d\theta^{2}
+\left(G_{R\theta}+G_{\theta R}\right)dR\,d\theta.
\nonumber
\end{align}
For the canonical complex scalar, the nonvanishing components are
\begin{equation}
G_{RR}=1,\qquad G_{\theta\theta}=R^{2},\qquad G_{R\theta}=G_{\theta R}=0.
\end{equation}

\subsection{Nontrivial internal field-space metric}\label{subsec:nontrivial_metric}

We now generalize to a complex scalar whose kinetic term defines a nontrivial
(in general curved and non-Euclidean) field-space metric, while keeping a
$U(1)$-symmetric potential $V(R)$ with no explicit $\theta$ dependence. As in
Eq.~\eqref{eq:canonical_field_metric}, we parameterize the internal line element
as
\begin{equation}
d\sigma^{2}=G_{IJ}(\phi)\,d\phi^{I}d\phi^{J}
=dR^{2}+f^{2}(R)\,d\theta^{2},
\label{eq:field_space_metric}
\end{equation}
where $f(R)$ is an arbitrary function of the radial field. For simplicity we
restrict to diagonal metrics ($G_{R\theta}=G_{\theta R}=0$). More general
non-diagonal metrics can introduce kinetic mixing between $R$ and $\theta$ and
lead to richer phenomenology. Absence of ghost instabilities requires a
positive-definite metric, in particular $f^{2}(R)>0$. The action becomes
\begin{align}
S=\int d^{4}x\,\sqrt{-g}\Big[&
-\frac{1}{2}g^{\mu\nu}\big(\partial_{\mu}R\,\partial_{\nu}R
+f^{2}(R)\,\partial_{\mu}\theta\,\partial_{\nu}\theta\big)
\nonumber\\
&-V(R)\Big],
\label{eq:general_action}
\end{align}
which reduces to the canonical flat-metric case when $f(R)=R$.

\subsection{Background equations of motion}\label{sec:background_eom}

We consider homogeneous background fields,
\begin{equation}
R=R(t),\qquad \theta=\theta(t),
\end{equation}
on an FRW background with $H\equiv\dot a/a$. The reduced action is
\begin{equation}
S_{\rm bg}=\int dt\,a^{3}(t)\left[\frac{1}{2}\dot{R}^{2}
+\frac{1}{2}f^{2}(R)\dot{\theta}^{2}-V(R)\right].
\label{eq:background_action}
\end{equation}
The background equations can be written covariantly as
\begin{equation}
D_{t}\dot{\phi}^{I}+3H\dot{\phi}^{I}+G^{IJ}V_{,J}=0,
\label{eq:covariant_eom}
\end{equation}
where $\phi^{I}=(R,\theta)$, $V_{,J}=\partial V/\partial\phi^{J}$, and $D_t$ is
the covariant time derivative associated with $G_{IJ}$. For the metric
\eqref{eq:field_space_metric}, the nonvanishing Christoffel symbols are
\begin{equation}
\Gamma_{\theta\theta}^{R}=-f(R)f'(R),
\qquad
\Gamma_{R\theta}^{\theta}=\Gamma_{\theta R}^{\theta}=\frac{f'(R)}{f(R)},
\label{eq:christoffels}
\end{equation}
with $f'(R)=df/dR$. The $R$ equation becomes
\begin{equation}
\ddot{R}+3H\dot{R}-f(R)f'(R)\,\dot{\theta}^{2}+V_{,R}(R)=0,
\label{eq:R_eom}
\end{equation}
and the $\theta$ equation is
\begin{equation}
\ddot{\theta}+2\frac{f'(R)}{f(R)}\dot{R}\,\dot{\theta}+3H\dot{\theta}=0.
\label{eq:theta_eom}
\end{equation}
For a $U(1)$-symmetric potential, Eq.~\eqref{eq:theta_eom} implies a conserved
charge,
\begin{equation}
a^{3}(t)\,f^{2}(R)\dot{\theta}=\text{const}.
\label{eq:charge_conservation}
\end{equation}
In this work we focus on the vanishing-charge sector,
\begin{equation}
\dot{\theta}_{0}=0,\qquad \theta_{0}=\text{const},
\end{equation}
for which the angular equation is automatically satisfied and the background
evolution is purely radial,
\begin{equation}
\ddot{R}_{0}+3H\dot{R}_{0}+V_{,R}(R_{0})=0.
\label{eq:R0_eom}
\end{equation}
Thus, in the zero-charge sector the homogeneous background does not depend
explicitly on the choice of $f(R)$.

\subsection{Fluctuations: radial $\delta R$ and canonical axial $\delta\psi$}
\label{subsec:fluctuations_zero_charge}

We now consider linear fluctuations about the homogeneous background,
\begin{equation}
R(t,\bm{x})=R_{0}(t)+\delta R(t,\bm{x}),
\qquad
\theta(t,\bm{x})=\theta_{0}+\delta\theta(t,\bm{x}),
\end{equation}
and work in a fixed FRW/de Sitter background (spatially flat gauge), neglecting
metric perturbations. Expanding Eq.~\eqref{eq:general_action} to quadratic order,
one finds that $\delta R$ and $\delta\theta$ decouple at linear order when
$\dot\theta_0=0$, with quadratic action
\begin{align}
S^{(2)}=\int d^{4}x\,a^{3}(t)\Bigg[
&\frac{1}{2}\left(\dot{\delta R}^{2}-\frac{(\nabla\delta R)^{2}}{a^{2}}
-V_{,RR}(R_{0})\,\delta R^{2}\right)
\nonumber\\
&+\frac{1}{2}f^{2}(R_{0})\left(\dot{\delta\theta}^{2}
-\frac{(\nabla\delta\theta)^{2}}{a^{2}}\right)\Bigg],
\label{eq:quadratic_action}
\end{align}
where derivatives of $V$ are evaluated on $R_0(t)$. Varying with respect to
$\delta R$ gives
\begin{equation}
\ddot{\delta R}+3H\dot{\delta R}-\frac{\nabla^{2}}{a^{2}}\delta R
+V_{,RR}(R_{0})\,\delta R=0.
\label{eq:dR_eom}
\end{equation}
The angular fluctuation is not canonically normalized. Defining the canonical
axial fluctuation
\begin{equation}
\delta\psi(t,\bm{x})\equiv f(R_{0}(t))\,\delta\theta(t,\bm{x}),
\label{eq:delta_psi_def}
\end{equation}
the $\delta\theta$ equation can be rewritten in Klein--Gordon form,
\begin{equation}
\ddot{\delta\psi}+3H\dot{\delta\psi}-\frac{\nabla^{2}}{a^{2}}\delta\psi
+m_{\psi}^{2}(t)\,\delta\psi=0,
\label{eq:dpsi_eom}
\end{equation}
with time-dependent effective mass
\begin{equation}
m_{\psi}^{2}(t)=\frac{f'(R_{0})}{f(R_{0})}V_{,R}(R_{0})
-\frac{f''(R_{0})}{f(R_{0})}\,\dot{R}_{0}^{2}.
\label{eq:mpsi2}
\end{equation}
Thus, in the vanishing-charge background, the internal geometry affects the
linear dynamics through the axial perturbations, while the radial perturbation
$\delta R$ is governed by the usual potential curvature $V_{,RR}(R_0)$.

\section{An example of hyperbolic PQ field-space\label{App:Hyperbolic_UV_example}}
In this Appendix, we illustrate an example of a hyperbolic manifold from a representative UV theory. While $\alpha$-attractors are traditionally used for inflaton potentials, we are repurposing this well-motivated geometry for a decoupled spectator sector \cite{Carrasco:2015rva,Carrasco:2015uma}). In our setup the PQ sector is described by a single complex scalar $\Phi$. At the level of a low-energy effective
theory, the most general two-derivative kinetic term takes the form of a
nonlinear sigma model,
\begin{equation}
\mathcal{L}_{\rm kin}
=\frac12\,g_{ab}(\phi)\,\partial_\mu\phi^{a}\partial^{\mu}\phi^{b},
\end{equation}
where $\phi^{a}$ are the two real components of $\Phi$. The PQ symmetry is
realized as a $U(1)$ isometry of the target space (shifts of the angular
coordinate). We now give a concrete example showing how a hyperbolic geometry can arise.

In $\mathcal{N}=1$ supergravity, scalar fields parameterize a K\"ahler manifold with metric (\cite{Wess:1992cp})
\begin{equation}
K_{i\bar{j}}=\frac{\partial^{2}K}{\partial \phi^{i}\,\partial\bar{\phi}^{\bar{j}}},
\end{equation}
and kinetic term
\begin{equation}
\mathcal{L}_{\rm kin}
=K_{i\bar{j}}\,\partial_\mu\phi^{i}\partial^{\mu}\bar{\phi}^{\bar{j}}.
\end{equation}
Appropriate choices of the K\"ahler potential $K$ generate hyperbolic metrics
already for a single complex field. A canonical example (familiar from
$\alpha$-attractor models \cite{Carrasco:2015rva,Carrasco:2015uma}) is the ``disk'' K\"ahler potential
\begin{equation}
K(Z,\bar Z)
=-\,3\alpha\,M_{P}^{2}\,
\ln\!\left(1-\frac{|Z|^{2}}{3M_{P}^{2}}\right),
\label{eq:diskK}
\end{equation}
which yields the $SU(1,1)/U(1)$-invariant K\"ahler geometry.

For \eqref{eq:diskK} one finds
\begin{equation}
K_{Z\bar Z}
=\frac{\alpha}{\bigl(1-|Z|^{2}/(3M_{P}^{2})\bigr)^{2}}.
\end{equation}
Parametrizing the complex field as $Z=\rho\,e^{i\theta}$, the kinetic term is
\begin{equation}
\mathcal{L}_{\rm kin}
=\frac{\alpha}{\bigl(1-\rho^{2}/(3M_{P}^{2})\bigr)^{2}}
\left[(\partial\rho)^{2}+\rho^{2}(\partial\theta)^{2}\right].
\label{eq:disk_kin_rhotheta}
\end{equation}
Matching to the sigma-model normalization
$\mathcal{L}_{\rm kin}=\frac12\,d\sigma^{2}$ gives
$g_{\rho\rho}=2K_{Z\bar Z}$ and $g_{\theta\theta}=2K_{Z\bar Z}\rho^{2}$.
Defining the canonically normalized radial field $R$ by
\begin{equation}
\frac{dR}{d\rho}
=\sqrt{2K_{Z\bar Z}}
=\frac{\sqrt{2\alpha}}{1-\rho^{2}/(3M_{P}^{2})},
\end{equation}
one obtains
\begin{equation}
R=\sqrt{6\alpha}\,M_{P}\,
\tanh^{-1}\!\left(\frac{\rho}{\sqrt{3}\,M_{P}}\right).
\end{equation}
Substituting back into \eqref{eq:disk_kin_rhotheta} yields the hyperbolic line
element
\begin{equation}
d\sigma^{2}
=dR^{2}+L^{2}\sinh^{2}\!\left(\frac{R}{L}\right)\,d\theta^{2},
\qquad
L=\sqrt{\frac{3\alpha}{2}}\,M_{P}.
\label{eq:hyperbolic_from_disk}
\end{equation}
Note that $R$ is unbounded, $R\in[0,\infty)$, so the regime $R\gg L$ is readily
accessible. The coordinate bound $\rho<\sqrt{3}\,M_{P}$ corresponds to the disk
boundary lying at infinite geodesic distance in field-space, i.e.\ it is a
coordinate artifact rather than a physical restriction.

We do not assume a complete supersymmetric completion in our analysis. However,
this example serves as a proof-of-principle and illustrates that a PQ-like complex scalar with hyperbolic field
space arises naturally in $\mathcal{N}=1$ supergravity, with the target-space
metric directly given by a well-motivated K\"ahler potential (and close cousins
appearing widely in inflationary model building). While this illustrative geometry captures the essential physics of isocurvature suppression, we acknowledge that a comprehensive UV completion-addressing the axion quality problem and higher-dimensional operator stabilization remains a distinct, significant task for future research.

\section{Semi-analytic template for the tilt and its running}
\label{app:template}

Here we sketch the derivation of the scale-space template used in the main
text. We work in the slow-roll geometric regime $\delta\equiv \xi(m_R/H)^2\lesssim 1$
(where $H\simeq H_{\rm inf}$ is approximately constant) and take the potential
to be quartic over the relevant field range.

\paragraph{Geometric regime.}
From Eq.~(\ref{eq:mpsi_hyp_2}) (or Eq.~(\ref{eq:mpsi_hyp}) in the $R\gtrsim L$
limit), the dominant contribution to the canonically normalized angular mass is
\begin{equation}
m_\psi^2 \simeq \xi\,m_R^2,
\qquad
\xi\equiv \frac{R}{L},
\qquad
m_R^2\equiv \frac{V_{,R}}{R},
\label{eq:geo_regime_app}
\end{equation}
where we neglect the subleading negative-curvature term $\propto-\dot R^2$ in the slow-roll limit.

\paragraph{Slow-roll background and $N$-scaling.}
For quartic slow-roll, $V_{,R}\simeq \lambda R^3$ and
\begin{equation}
3H\dot R \simeq -V_{,R}\simeq -\lambda R^3.
\end{equation}
In terms of efolds $N\equiv \ln a$ (so $dN=Hdt$), this gives
\begin{equation}
\frac{dR}{dN}\simeq -\frac{\lambda}{3H^2}\,R^3
\quad\Rightarrow\quad
\frac{1}{R^2(N)}=\frac{1}{R_p^2}+\frac{2\lambda}{3H^2}\,(N-N_p),
\label{eq:R_of_N_app}
\end{equation}
where $N_p$ is a reference efold at which we match onto the geometric regime and
$R_p\equiv R(N_p)$. Using $m_R^2=\lambda R^2$ and $\xi=R/L$, Eq.~(\ref{eq:geo_regime_app})
implies
\begin{equation}
m_\psi^2(N)\simeq \frac{\lambda}{L}\,R^3(N).\label{eq:mpsi_R3_app}
\end{equation}
Combining with Eq.~(\ref{eq:R_of_N_app}) yields the characteristic falloff
\begin{equation}
m_\psi^2(N)\simeq m_{\psi,p}^2\left[
1+\frac{2}{3}\left(\frac{m_{\psi,p}^2}{H^2\xi_p}\right)(N-N_p)
\right]^{-3/2},
\label{eq:mpsi_of_N_app}
\end{equation}
where $m_{\psi,p}^2\equiv m_\psi^2(N_p)$ and $\xi_p\equiv R_p/L$.
(The identity used above is $m_{\psi,p}^2/H^2=\xi_p\,m_{R,p}^2/H^2$.)

\paragraph{$k$-space template.}
For nearly constant $H$, horizon exit implies $k=aH$ and hence
\begin{equation}
\ln\!\left(\frac{k}{k_p}\right)\simeq N-N_p,
\label{eq:kNmap_app}
\end{equation}
with $k_p\equiv a(N_p)H$. Substituting Eq.~(\ref{eq:kNmap_app}) into
Eq.~(\ref{eq:mpsi_of_N_app}) gives Eq.~(\ref{eq:mpsik_template}) in the main text,
\begin{equation}
m_{\psi}^{2}(k)\simeq m_{\psi}^{2}(k_p)\left[
1+\frac{2}{3}\left(\frac{m_{\psi}^{2}(k_p)}{H^{2}\xi_p}\right)
\ln\!\left(\frac{k}{k_p}\right)\right]^{-3/2}.
\end{equation}
Together with the massive-mode estimate \eqref{eq:niso}, this yields a unique analytic
template for the tilt and its running for modes exiting after matching,
$k\gtrsim k_p$.

In the geometric slow-roll regime one obtains for $k\gtrsim k_p$,

\begin{equation}
m_\psi^2(N)\propto (N-N_p)^{-3/2}
\,\,\leftrightarrow\,\,
m_\psi^2(k)\propto \left[\ln\!\left(\frac{k}{k_p}\right)\right]^{-3/2},
\end{equation}which is the origin of the characteristic non-power-law running emphasized in
the main text.

%\nocite{apsrev%41Control}
% \bibliographystyle{apsrev4-1}
\bibliography{ref}

\end{document}